\documentstyle[twoside,fleqn,rgs_workshop,psfig]{article}

\begin{document}

\title{\huge{Ratio of color and effective temperatures in X-ray burst
spectra}}

\author{A. Majczyna  
\address{Copernicus Astronomical Center, Bartycka 18, PL-00-716 Warsaw, Poland
(majczyna@camk.edu.pl)}
J. Madej, 
\address{Astronomical Observatory, University of Warsaw, Al. Ujazdowskie 4,
PL-00-478 Warsaw, Poland (jm@astrouw.edu.pl)},
P.C. Joss 
\address{Department of Physics, Center for Space Research, and Center for
Theoretical Physics, Massachusetts  Institute of Technology, Cambridge,
Massachusettss 02139 (joss@mitlns.mit.edu)},
A. R\'o\.za\'nska  
\address{Copernicus Astronomical Center, Bartycka 18, PL-00-716 Warsaw, Poland
(agata@camk.edu.pl)}
}
 
\begin{abstract}
We present model atmospheres and theoretical spectra for X-ray
bursters.  Our models include the effects of Compton scattering on free
electrons.  The atmospheres have compositions that are mixtures of
hydrogen, helium, and iron.  For our models the ratio of color
temperature, $T_c$, to effective temperature, $T_{eff}$, is in the range 1.2 -
1.8.  This ratio depends on $T_{eff}$, surface gravity, and iron abundance.
For fixed $T_{eff}$ and surface gravity, models with non-zero iron abundance
exhibit values of $T_c/T_{eff}$ that are slightly lower than pure
hydrogen-helium models.

\end{abstract}

\maketitle

\section{Introduction}
Discovery of the first X-ray burster was reported by Grindlay et al.
(1976)\cite{grindlay} and by 
Belian et al. (1976)\cite{belian}. We believe that these objects are low mass
binary stars
which contain the accreting neutron star. Since we do not observe X-ray
pulsations, 
the magnetic field of the neutron star is lower than $10^{10}$ Gs. Weak magnetic
field of such intensity allows for the existence of accretion disc. Basic
features of X-ray
burst are: fast rise $\sim$ 1 s and decay time $\sim$ 3 -- 100 s, luminosity
near the 
peak about $10^{38}$ erg/s and energy per burst $~10^{39}$ ergs (cf. review 
article by Joss \& Rappaport 1984\cite{joss}). The idea that energy source of a
burst is the 
thermonuclear flash was proposed by Woosley and Taam (1976)\cite{woosley} and
Maraschi \& 
Cavaliere (1977)\cite{maraschi}. This model quite well explains many global
features of X-ray 
bursters. 

Initially neutron star spectra were assumed to be blackbody spectra,
but this assumption is not valid when the atmosphere was dominated by
scattering on free electrons. In such case spectra of outgoing
photons are shifted towards
higher energy and the maximum is shifted in the same direction both for the
coherent 
Thomson, and noncoherent Compton scattering. In electron scattering stellar
atmosphere, photons are created by
thermal processes below the photosphere, in so called thermalization
layer. In this layer local temperature T is much
higher
than the effective temperature $T_{eff}$, therefore maximum of the Planck
function $B_{\nu}(T)$  occurs in higher energy than the maximum of
$B_{\nu}(T_{eff})$. Consequently, the blackbody temperature $T_{BB}$
as obtained from fitting of continuum spectra, can be much larger than the
true $T_{eff}$. Temperature $T_{BB}$ is usually 
referred to as the color temperature $T_c$.

In principle, the ratio $T_{c}/T_{eff}$ can reach arbitrary large values in
Thomson scattering stellar atmospheres, provided that true absorption in
that atmospheres decreases to zero (Madej 1974 \cite{madej74}). Compton
scattering in stellar atmospheres reduces  $T_{c}/T_{eff}$ by unknown
factor, but still this ratio remains larger than unity. Model atmosphere
computations with account of Compton scattering, and exact $T_{c}/T_{eff}$
determination,  are very difficult.

Ebisuzaki, Hanawa \&
Sugimoto (1984)\cite{ehs} estimated $T_c/T_{eff}\approx1.4$ by use the simple
opacity
formulae. Model atmospheres computations by London, Taam \& Howard
(1986)\cite{lth}
showed,
that this ratio was in the range from 1.2 to 1.7 and depended on few parameters:
$T_{eff}$, surface gravity log g and iron abundance. London, Taam \& Howard
(1986) \cite{lth}
concluded that the solar iron abundance does not influence significantly the
ratio
$T_{c}/T_{eff}$. On the other hand, Ebisuzaki \& Nomoto (1986)\cite{ebisuzaki}
claimed that
absorption by iron decrease this ratio.

\section{Model equations}
Our model equations describe transfer of radiation subject to constraints of
hydrostatic
and radiative equilibrium. We assume the equation of state of ideal gas in local
thermodynamical equilibrium (LTE). The equation of transfer includes f-f
absorption from
all
ions, b-f absorption from iron ions, and Compton scattering terms. The equation
of
transfer is as follows: 
\begin{eqnarray}
\mu\frac{\partial
I_{\nu}}{\partial\tau_{\nu}}&=&I_{\nu}-\epsilon_{\nu}B_{\nu}-(1-\epsilon_{\nu})J_{\nu}
\nonumber  \\
&+&(1-\epsilon_{\nu})J_{\nu}\int_0^{\infty}
\Phi_1(\nu,\nu')d\nu'
\nonumber \\
&-&(1-\epsilon_{\nu})\int_0
^{\infty} J_{\nu'}\Phi_2(\nu,\nu')d\nu' \nonumber
\end{eqnarray}
where: $B_{\nu}$ is the Planck function, $J_{\nu}$ the mean intensity,
$\epsilon_{\nu}$
is the ratio of true absorption to coefficient of total opacity, $\tau_{\nu}$
the
monochromatic optical depth, and $\mu$
denotes cosine of zenithal angle.\\
Details of the above equation of transfer, and Compton scattering redistribution
functions $\Phi_1(\nu,\nu')$ and $\Phi_2(\nu,\nu')$, and linearization
procedure,  are described in earlier
papers  (see Madej 1991 \cite{madej}; Madej \& R\'o\.za\'nska 2000a\cite{mra},
b\cite{mrb}; Madej, Joss \&
R\'o\.za\'nska 2003\cite{mjr2003}).     

\section{Computations}
We calculated our models with the following assumptions: equation of state in
LTE, 
hydrostatic and radiative equilibrium, plane-parallel geometry, non rotating
star.
Our equations allow for large photon-electron energy exchange at the time of
Compton
scattering.  Details of the computer code and linearization scheme were
extensively
described in earlier papers quoted in previous section. 

We calculated grid of models each of them computed on the 144 standard optical
depth
points (12 points per the decade)
from $10^4$ to $10^{-8}$ and 901 wavelength points. We computed few models of
hot neutron star atmospheres with effective temperatures $T_{eff}$=1, 2,
3$\times10^7$ K, and surface gravities $g$ from the critical gravity up to
$10^{15}$
in 
cgs units. Models assume H and He in solar number abundance $N_{He}/N_H=0.11$
and iron
with number abundance $N_{Fe}/N_H=3.7\times10^{-5}$ (approximately solar value)
and 100 times
greater.  

\section{Results}
We claim that our computations can measure both the ratio $T_c/T_{eff}$ and its
slight changes influenced by small differences of $T_{eff}$, $\log g$ and iron
abundance. We obtain models of burster
atmospheres where the constancy of the bolometric flux is almost everywhere
better than 1\%, and only in few optical depth points approches 2\%.

Tables 1 and 2 show values of $T_c/T_{eff}$ computed over wide range of
parameters
$T_{eff}$, $\log g$ and
iron abundance. We can see that for models with pure H-He atmospheres, these
ratios are
higher than for atmospheres with nonzero iron abundance, of course for the same
$T_{eff}$ and $\log g$. For higher metal abundance these ratios decrease. 
This is because increase of metal abundance diminish the importance of
Compton scattering, and $T_c/T_{eff}$ ratio is than determined by the
redistribution of outgoing flux by nongray metal absorption. 

\begin{table}[!h]
\footnotesize{
\caption{$T_c/T_{eff}$ ratios for different parameters. \newline Left part of
the tables present models with $N_{He}/N_H=0.11$ and $N_{Fe}/N_{H}=10^{-5}$,
and right part $N_{He}/N_H=0.11$, and $N_{Fe}/N_{H}=10^{-3}$ }
  $$
\begin{array}{||c|c|c|||c|c|c||} 
\hline \hline
T_{eff} & \log g &  T_{c}/T_{eff}&T_{eff} & \log g &  T_{c}/T_{eff} \\ \hline 
            & 14.8 &  1.77 &    &14.8 &  1.25  \\
3\times 10^7 & 14.9 &  1.56 &  3\times 10^7 &14.9 & 1.24 \\
            & 15.0 &  1.44 &              &15.0 & 1.24\\ \hline
            & 14.1 &  1.67 &      &14.0 & 1.41\\
2\times 10^7 & 14.5 &  1.41 &  2\times 10^7   &14.5 & 1.38 \\
            & 15.0 &  1.27 &      &15.0 & 1.27 \\ \hline
            & 12.9 &  1.51 &    &12.8 & 1.27\\
 1\times 10^7& 14.5 &  1.22 &  1\times 10^7  &14.5 & 1.30\\
           & 15.0 &  1.25 &            &15.0 & 1.36 \\ \hline \hline
\end{array}
   $$}
\end{table}
\begin{table}[!h]
\small{
\caption{Table shows ratios $T_c/T_{eff}$ for pure H-He model atmospheres
(Madej, Joss \& R\'o\.za\'nska 2003)} 
\vskip3mm
 $$
\begin{array}{||c|c|c||}
\hline \hline
T_{eff} & \log g &  T_{c}/T_{eff} \\ \hline
        & 14.8 &  1.79   \\
3\times 10^7 & 14.9 &  1.69\\
            & 15.0 &  1.61 \\ \hline 
            & 14.1 &  1.75 \\
2\times 10^7 & 14.5 &  1.43 \\
             & 15.0 &  1.29 \\ \hline 
             & 12.9 &  1.75 \\
1\times 10^7 & 14.5 &  1.29 \\
            & 15.0 &  1.33 \\ \hline \hline
\end{array}  
  $$}
\end{table}

What is interesting, for the lowest $T_{eff}=10^7$K
these ratios exhibit local mimimum in some intermediate $\log g$. Tendencies,
which we observe are due to iron opacity and the Compton scattering. For the
lowest surface gravity 
Compton scattering dominates over thermal absorption,  therefore  shift of the
peak
spectra is the highest. In the presence of iron, thermal absorption is
greater than in
case of H-He atmospheres.
Consequently, the $T_c/T_{eff}$ is lower for atmospheres containing iron.

In Figure 1 we show the outgoing photon spectra for various parameters. We
compare blackbody spectra with the spectra of H-He atmosphere and with the
atmosphere
containing iron in Figure 1a. We can see that the blackbody spectra (solid line)
have the peak in
lower energy and the BB spectrum is softer than the spectrum of the model X-ray
burster.
In lower photon energies we obtain excess of the outgoing flux as compared to
the blackbody spectra both for H-He models and atmospheres containing some
amount of iron. Figures 1c and 1d present spectra of
atmospheres with iron abundance 100 times greater
than the solar abundance, and with the solar iron abundance, respectively. For
the latter iron abundance we note, that
the iron ionisation edge appear in emission for two lowest
$T_{eff}$. Profiles of iron absorption lines (fundamental series of H-like,
and He-like ions) can be very complicated (see Fig. 1b). In some models we
obtain inverse
emission in iron line cores, which strictly correspond to temperature inversion
reproduced in our model atmospheres with Compton scattering.

We stress here, that we did not obtain values of $T_c/T_{eff}$ larger than
2.0 in any of our model atmospheres of X-ray burst sources.

\section*{ACKNOWLEDGEMENTS}
We acknowledge support of the grant No. 2 P03D 021 22 from the Polish
Committee for Scientific Research.

\begin{figure*}[h] 
\vspace{10pt}
{\psfig{file=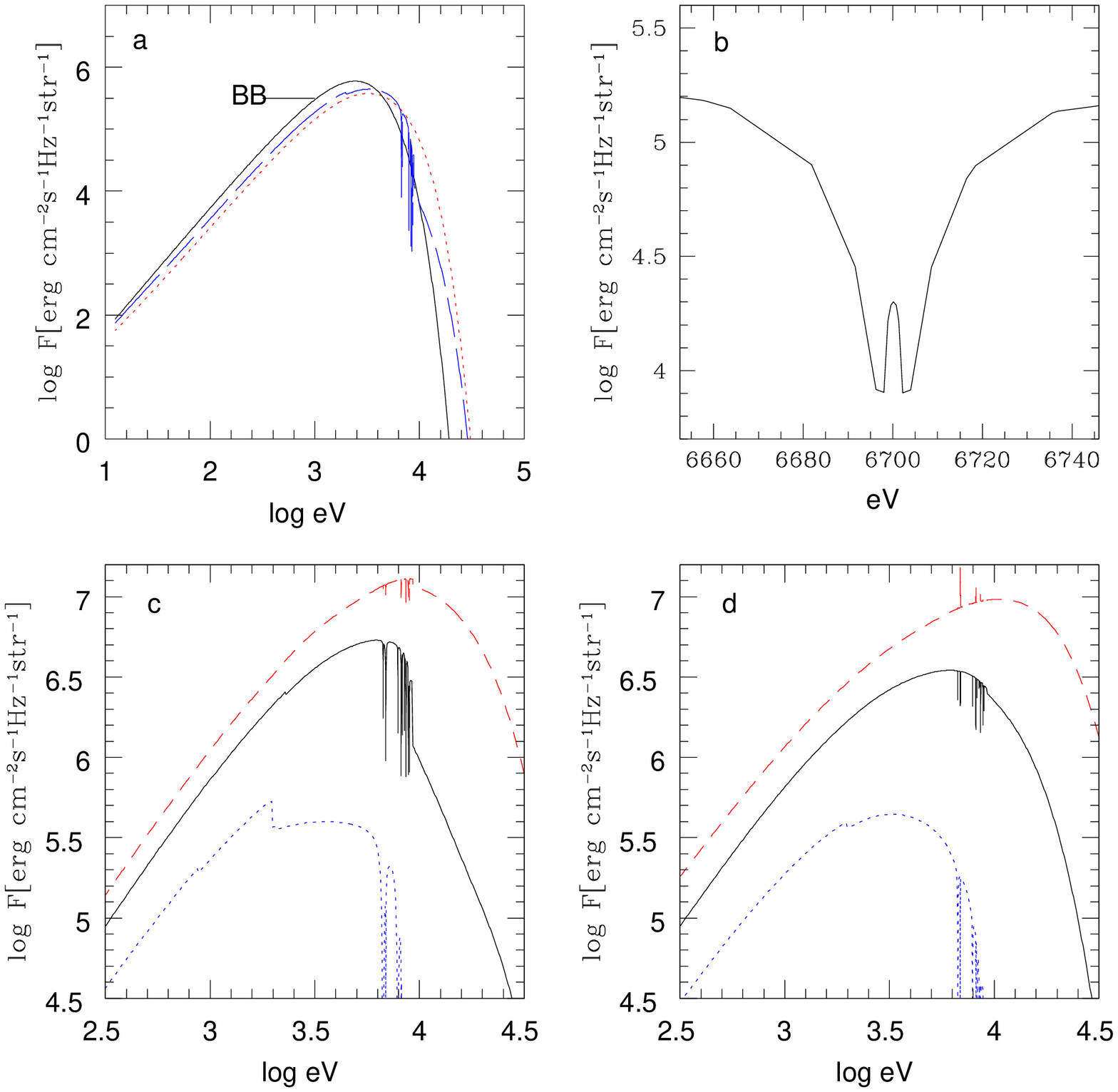,width=6.0in,height=4.0in,angle=0}}
\caption{Various models of neutron star atmospheres.\newline
a) solid line - blackbody spectrum with $T=10^7$ K; dotted line - pure H-He
$\log g=15$, $T_{eff}=10^7$ K; dashed line $N_{Fe}/N_H=3.7\times10^{-5}$, $\log
g=15$, $T_{eff}=10^7$ K;\newline
b) resonant line profile of helium-like iron;\newline
c) $N_{Fe}/N_H=3.7\times10^{-3}$
dotted  line $\log g=15$, $T_{eff}=10^7$ K; solid $\log g=15$,
$T_{eff}=2\times10^7$ K; dashed $\log g=15$, $T_{eff}=3\times10^7$ K; \newline
d) $N_{Fe}/N_H=3.7\times10^{-5}$
dotted  line $\log g=15$, $T_{eff}=10^7$ K; solid  $\log g=15$,
$T_{eff}=2\times10^7$ K; dashed $\log g=15$, $T_{eff}=3\times10^7$ K. }
\end{figure*}
\end{document}